# A mathematical theory of power

Author: Daniele De Luca, University of Turin
E-mail: d.deluca@unito.it
Postal address: Via XXVII Marzo, 10093, Collegno (TO), Italy

*Abstract*

This paper proposes a new approach to power in Game Theory. Cooperation and conflict are simulated with a mechanism of payoff alteration, called F-game. Using convex combinations of preferences, an F-game can measure players' attitude to cooperate. We can then define actual and potential power as special relations between different states of the system.

## *1.1 Introduction*

Power is a social relation among individuals that affects their behaviors. In order to develop a mathematical theory of power, we need a framework of actors with predictable behavior that power can affect. Game theory is a good candidate: there are players that behave consistently with their preferences, and the action of a player can alter the welfare of the other.

Shapley and Shubik (1954), followed by many others, e.g. Allingham (1975) and Wiese (2009), have proposed indices of power in the context of Cooperative Game Theory. The Shapley value and the Banzhaf index (see Banzhaf (1964)) measure power as a special relation among players and coalitions. They became a basic concept for evaluating power in voting games. But they are all dependent on the same type of coalition among players, where players can only be in or out. Cooperative Game Theory excludes individual actions to concentrate only on groups and their possible collective payoff, and this seems a limit in the analysis of social relations.

Power has also been described with formal methods outside game theory, in the context of sociology. So called Dependence Theory developed several tools for mathematically expressing concepts like resource or social dependence and social influence, for example in Castelfranchi et al. (1992), Castelfranchi (1998), Sichman (1998) and Boella et al. (2005).

Eventually, Grossi and Turrini (2010, 2012) proposed a way to apply Dependence Theory on Game Theory. They used coalitional games and Shapley value to develop a general measure of dependence in games.

This paper proposes a new approach, yet not incompatible with the formers: I introduce cooperation and conflict with a mechanism of payoff alteration, called $F$-game. An $F$-game can not only simulate coalitions, it also measures players' attitude to cooperate. Furthermore, it includes a new possibility: players can act against other players.

But the main feature that distinguishes an $F$-game from a coalitional game is that the former includes individual preferences. While cooperative games substitute the whole non-cooperative structure with a characteristic function, $F$-games maintain it as a subset.

*1.2 General definitions of power*

When dealing with the concept of power, it is very important to avoid ambiguity. There is more than one meaning of the word power, as it has been widely stressed. For instance, one can find a huge amount of literature on the distinction between *power over* and *power to*, or between *potential* and *actual power*. So, before giving a mathematical shape to power, it is worth clearly defining the words we are going to use.

Let us begin with some general definitions.

In his influential essay *The Concept of Power,* Robert Dahl (1957: 202-3) gave this famous definition: "A has power over B to the extent that he can get B to do something that B would not otherwise do." Michael Taylor (1982: 13) seems to follow Dahl's assertion when he says that power is "the ability to affect the incentives facing others so that it becomes rational for them to pursue a certain course of action." According to Pranab Bardhan (1991: 274), in terms of game theory, "an inclusive way of defining power may be to say that A has power over B if A has the ability to alter the game (preferences, strategies, sets or sets of information) in such a way that B's equilibrium outcome changes." Randall Bartlett (2006: 30) says, in turn, that power is "the ability of one actor to alter the decisions made and/or welfare experienced by another actor relative to the choices that would have been made and/or welfare that would have been experienced had the first actor not existed or acted."

All these definitions may agree with the idea that there is a change in the behavior of the submitted player. What they lack is to specify the *direction* of the changes. It may not be bold to think that the direction is the welfare of the dominant. Samuel Bowles' (1998: 11) asymmetric definition takes into account also aspect: "For agent A to have power over agent B it is sufficient that, by imposing or threatening to impose sanctions on B, A is capable of affecting B's actions in ways which further A's interests, while B lacks this capacity with respect to A." The same for Mario Stoppino's definition (in Matteucci et al. 1991): "The behavior of A, who exercises power, can be associated, rather than with the intention of determining the behavior of B, the object of power, with the interest that A has in such behavior."

The previous definitions contain two main parts: when A has power over B, there is a capacity of A of doing something (normally referred to by the verb "to can", or by some "ability" to impose some incentives, threats, and so on) and a change in the behavior of B. This change can be further specified as consistent with A's interests.

A's ability of making promises or threats is what I will call *potential power*. On the other hand, the change in B's behavior (or welfare) is what I will call *actual power* or *influence/sympathy*. It is clear from the previous definitions that these two phenomena are quite often linked, although their link is not deterministic nor necessary.

It is normally assumed that actors' preferences are fixed and cannot change without changing the whole game. When we change the game, we are supposed to do some mechanism design. However, all these definitions suggest that the concept of power can only be expressed through some alteration of welfare, preferences, and strategies. Therefore, we have to deal with some modifiers of the preference functions.

## *2.1 Actual power definition*

Although actual power has been defined as a kind of favoring, it seems natural to include the possibility of harming. Besides, this will be necessary in order to further develop the notion of potential power, as we will see in section 4.

DEFINITION: If B acts for the sake of A, I say that A *positively influences* B, or that B has *positive sympathy* for A. On the converse, when B acts against A, I say that A *negatively influences* B, or that B has *negative sympathy* for A. I refer to this type of behavior in general as *actual power*.

The easiest way to represent a change in B's behavior is altering B's payoffs. Therefore, I will introduce the notion of *mixed preference*, as opposed to *pure preferences*, in analogy with the common distinction between *pure* and *mixed strategies*. From a mathematical standpoint, a *mixed strategy* for player B is a convex combination of B's *pure strategies*. In quite the same way, a *mixed preference* for player B is a convex combination of all players' *pure preferences*, including B's. With mixed preferences we can represent the behavior of favoring or harming another.

There is abundant literature on the subject of so-called "social preferences" that claims that human behavior can be influenced by social values such as fairness, equity and so on. While no one denies that possibility, the very weight of social preferences in human behavior is a controversial topic. See for instance Ernst Fehr and Gary Charness (2023) for an overall review and Ken Binmore (2010) and John A. List (2009) for more critical accounts of the field. This paper does not consider universal preferences such as fairness, but only individual-specific preferences such as someone's willingness to favor or harm a particular player.

The possibility of mixing players' goals is implicit in the commonly accepted definition of rationality as *rationality of means* and of preferences as *revealed preferences*. Ken Binmore

(2009) states that "there can then be nothing irrational about consistently pursuing any end whatever. As Hume extravagantly observed, he might be criticized on many grounds if he were to prefer the destruction of the entire universe to scratching his finger, but his preference could not properly be called irrational, because (contra Kant) rationality is about means rather than ends." Similarly, the idea behind the revealed preferences is that there is nothing *a priori* in the preferences of an agent, so that we can determine them only *a posteriori*, when they reveal in the agent's actions. The conclusion is that an agent could have any kind of preference, even an altruistic or masochistic one. In particular, agents can aim to favor or harm each other.

## 2.2 Actual power in games

Let us introduce the notion of $F$-games. An $F$-game is described by the following tuple.

**D1.**  $\Gamma = (N, S_{i \in N}, u_{i \in N}, F)$

$N$ is the set of players.

$S_{i \in N}$ are the sets of possible strategies for each player $i$. I use uppercase $S_i$ to denote the entire (discrete or continuous) set of strategies of a player $i$, lowercase $s_i$ to instantiate one of these strategies (or a value if $S_i$ is continuous).

$u_i$ are the *pure preference* (or *pure utility*) functions. For any profile $(s_1, s_2, \ldots, s_n)$, where $s_i$ represents the strategy chosen by player $i$, the function $u_i(s_1, s_2, \ldots, s_n)$ returns a single value that represents $i$'s pure preference for that profile, when no other player influences $i$.

Note that there is no absolute value of a profile for a player $i$, but rather a relative one. It is only the difference of preference between two strategies that informs us which will be chosen. From a mathematical perspective, a function $u_i$ is equivalent to any function $u_i + c$, where $c \in R$ is any real number. Nevertheless, in $F$-games preference functions are not identical up to multiplication for a positive constant, because they require some form of comparison of the different utilities among players.

$F \in R^{n,n}$ is an adjacency matrix (or a graph) that represents the influence, be it positive or negative, that each player has toward another. In other words, $F$ represents actual power. I take 1 as the limit influence that a player can be affected, so:

**A1.** $\quad \sum_{j=1}^{n} |f_{j,i}| < 1$

Furthermore, the diagonal of $F$ will always be null ($f_{i,i} = 0$), as self-loops are not allowed.

If $F$ is null, the game works in the traditional way. Otherwise, if any entry of $F$ is not null, some or all the players' choices may change. With only two players, when $i$ is influenced by $j$ by the factor $f_{j,i}$, we define $U_i$ as the *mixed preference* (or *mixed utility*) of $i$ in the following way:

**F1.** $\quad U_i = f_{j,i} U_j + (1 - |f_{j,i}|) u_i$

I intend that the players always choose the strategies that maximize $U_i$, the *mixed preference*, and not the ones that maximize $u_i$, the *pure preference*, unless they are free from influence. For this reason, it is necessary that players influenced by $j$ approximate their utility to $U_j$, which represents the actual preference of that player, and not to $u_j$. In this way, we can take into account simultaneity of influences and their possible circularities.

As we see in F1, $f_{j,i}$ is the linear measure of how much the mixed preference of $i$ resembles the mixed preference of $j$. If $f_{j,i}$ is 0, there is no influence and $U_i = u_i$. This means that player $i$ is free from influence. If $f_{j,i}$ is near 1, there is big positive influence and $U_i \approx U_j$. Eventually, if $f_{j,i}$ is near -1, there is big negative influence and $U_i \approx - U_j$.

Let us generalize the formula for $n$ players:

**D2.** $\quad U_i = \sum_{k=1}^{n} f_{k,i} U_k + (1 - \sum_{y=1}^{n} |f_{y,i}|) u_i$

To compute each mixed utility given the known pure ones, we need to transform the matrix $F \in R^{n,n}$ into a new matrix $C \in R^{n,n}$. In what follows I will call the entries of $C$ "colonizations" to distinguish them from the "influences" of $F$. "Colonization" refers to the ultimate influence of a player on another once every edge has been covered. The aim is to calculate each mixed $U_i$ function as a special convex combination of pure $u_i$ functions, as follows:

**D3.** $$U_i = \sum_{j=1}^{n} c_{j,i} u_j$$

Substituting D3 in D2 we now obtain:

**F2.** $$\sum_{j=1}^{n} c_{j,i} u_j = \sum_{j=1}^{n} \sum_{k=1}^{n} f_{k,i} c_{j,k} u_j + \left(1 - \sum_{j=1}^{n} |f_{j,i}|\right) u_i$$

The following two formulas F3 and F4 allow us to build up a linear system of $n^2$ equations in $n^2$ variables and find the partial values $c_{j,i}^p$. Both are derived from F2, considering the set $\{u_1, u_2, \dots, u_n\}$ as the basis of a vector space.

**F3.** $$c_{j,i}^p = \sum_{k=1}^{n} f_{k,i} c_{j,k}^p \qquad (j \neq i)$$

**F4.** $$c_{i,i}^p = \sum_{k=1}^{n} f_{k,i} c_{i,k}^p + 1 - \sum_{j=1}^{n} |f_{j,i}|$$

Eventually, with F5 we normalize the partial colonizations into constant sum ones, in order to verify F6 and obtain the convex combination of D3.

**F5.** $$c_{j,i} = \frac{c_{j,i}^p}{\sum_{k=1}^{n} |c_{k,i}^p|}$$

**F6.** $$\sum_{j=1}^{n} |c_{j,i}| = 1$$

The value $c_{i,j}$ in the $i$-th row and in the $j$-th column indicates the colonization of $i$ in $j$. If $c_{i,j} > 0$ then we say that $i$ *colonizes* $j$. In other words, $c_{i,j}$ says how much $U_j$ is affected by $u_i$, while $f_{i,j}$ says how much it is affected by $U_i$.

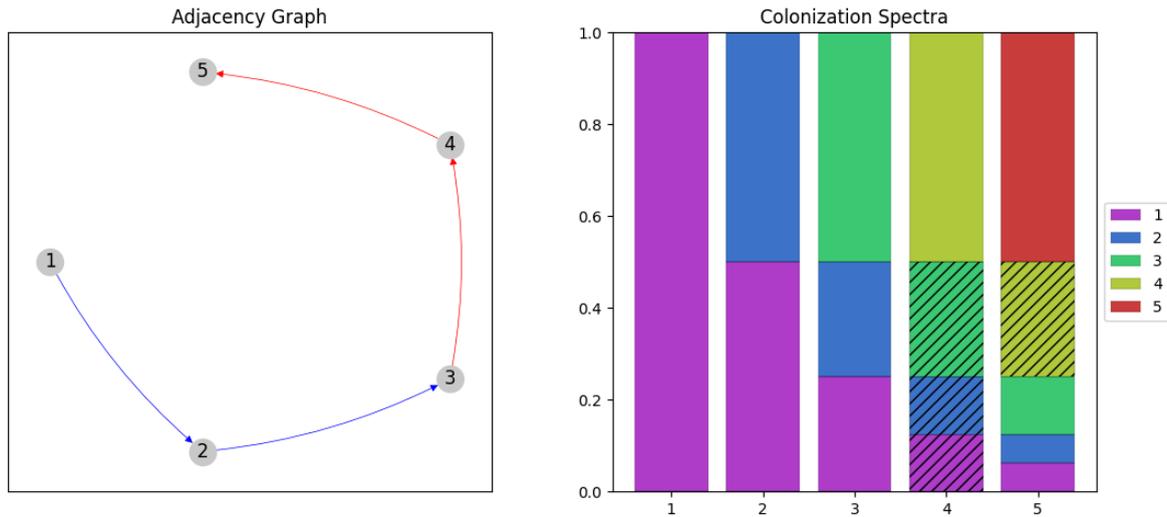

**Figure 1.** *A representation of an F-matrix as a graph (left) and a C-matrix as a histogram (right).*

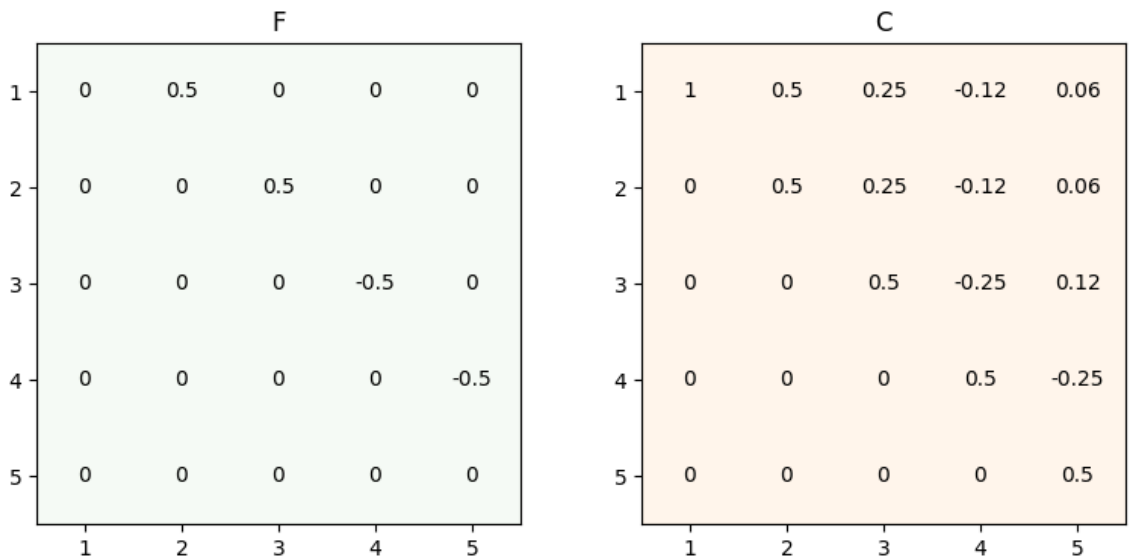

**Figure 2.** *F-matrix (left) and C-matrix (right).*

The left side of *Figure 1* is a representation of the matrix $F$ in *Figure 2*. The edges are blue where positive, red where negative.

The right side of *Figure 1* is a representation of the relative matrix $C$, shown in *Figure 2*, in the form of a histogram. It is worth noting how colonizations are transitive. The diagonally barred colors are the negative colonizations.

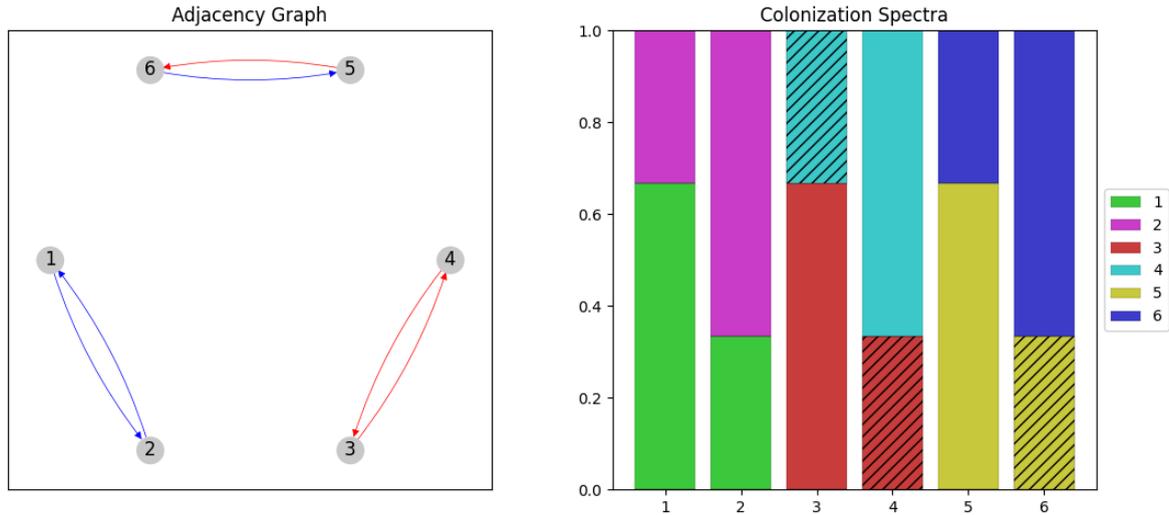

**Figure 3.** *Reciprocity in F and C (graph and histogram).*

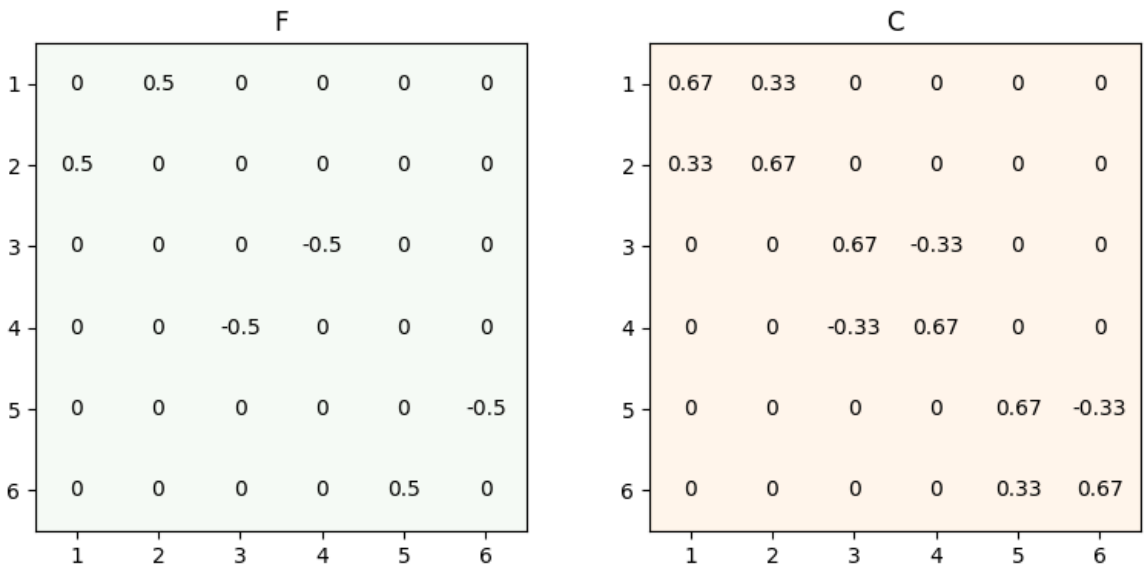

**Figure 4.** *Reciprocity in F and C (matrices).*

Figure 3 shows some examples of circularity. Two player cycles can be positive, negative, or positive-negative. In a cycle, colonization is normally less than influence. In *Figure 4*, for example, $f_{1,2} = \frac{1}{2}$, while $c_{1,2} = \frac{1}{3}$.

We now introduce a definition of Nash equilibrium in $F$-games. A profile $\sigma = (s_1, s_2, \ldots, s_n)$ is an *F-equilibrium* if, for each player $i$, $s_i$ is the best response to the strategies of the other $n-1$ players, according to $i$'s mixed utility function. D4 is a formalization of this concept, with $\sigma = (s_1, \ldots, s_i, \ldots, s_n)$ and $\sigma[s'_i] = (s_1, \ldots, s'_i, \ldots, s_n)$.

**D4.** σ is *F*-equilibrium iff $\forall i \in N : \forall s'_i \in S_i : U_i(\sigma) \geq U_i(\sigma[s'_i])$

Therefore, in an *F*-game any equilibrium depends on *F*, because the $U_i$ functions can change when *F* changes. For each profile $\sigma = (s_1, s_2, \ldots, s_n)$ we can ask for what values of *F*, if there are any, σ is an equilibrium. We define those values the "influence space" $F_\sigma$ of a profile σ:

**D5.** $F_\sigma = \{F \in R^{n,n} : \forall i \in N : \forall s'_i \in S_i : U_i(\sigma) \geq U_i(\sigma[s'_i])\}$

If the set $F_\sigma$ includes the null matrix, σ is a Nash equilibrium in the traditional sense. But not every Nash equilibrium has the same influence space. In strategic form games we can often classify Nash equilibria via their influence spaces.

To check if a value of *F* is compatible with a profile σ, we substitute D3 in D5 to find $C_\sigma$, the colonization space of σ:

**F7.** $C_\sigma = \{C \in R^{n,n} : \forall i \in N : \forall s'_i \in S_i : \sum_{j=1}^{n} c_{j,i} u_j(\sigma) \geq \sum_{j=1}^{n} c_{j,i} u_j(\sigma[s'_i])\}$

*2.3 Influence space of a strategy profile*

Let us restrict the game to only two players, 1 and 2, with only two choices per player. We define $a$ as the difference between the preferences of player 1 for σ and for $\sigma[s'_1]$:

**D6.** $a = u_1(\sigma) - u_1(\sigma[s'_1])$

Then, we define $b$ as the same difference for player 2:

**D7.** $b = u_2(\sigma) - u_2(\sigma[s'_1])$

From F7, we obtain that the colonization $c_{2,1}$ has to obey one of the following inequalities to make σ an F-equilibrium. For every $s'_1 \neq s_1$:

**F8.** $\quad c_{2,1} \geq \dfrac{-a}{b+sign(b)|a|} \quad$ if $sign(b) = 1$

$\quad\quad c_{2,1} \leq \dfrac{-a}{b+sign(b)|a|} \quad$ if $sign(b) = -1$

In the same way we find the inequalities for $c_{1,2}$, switching between the two players in D6, D7 and F8. These two inequalities describe the colonization space of σ, or $C_\sigma$.

### 3.1 The Prisoner's Dilemma

Let us take as an example a typical Prisoner's Dilemma.

**PRISONER'S DILEMMA**          Player 2
(*Table 1*)

|  |  | L | R |
|---|---|---|---|
| Player 1 | U | -1, -1 | -6, 0 |
|  | D | 0, -6 | -5, -5 |

We want to find the influence space of $DR$, the Nash equilibrium of the game. We apply F8 for the two players and we find:

**F9.** $\quad C_{DR} = \{\forall C : c_{2,1} \leq \frac{1}{6} \land c_{1,2} \leq \frac{1}{6}\}$

This is the colonization space of $DR$, depicted in *Figure 5* as a green area. The black square marks the limiting condition $|c_{2,1}| + |c_{1,2}| < 1$ that can be obtained from A1, F4 and F6. The centroid of the green polygon is $H = (-\frac{4}{15}, -\frac{4}{15})$. This measure can be used to classify profiles and Nash equilibria in strategic games. The blue region is the colonization space of $UL$, while the two white regions are, from the left, $DL$ and $UR$.

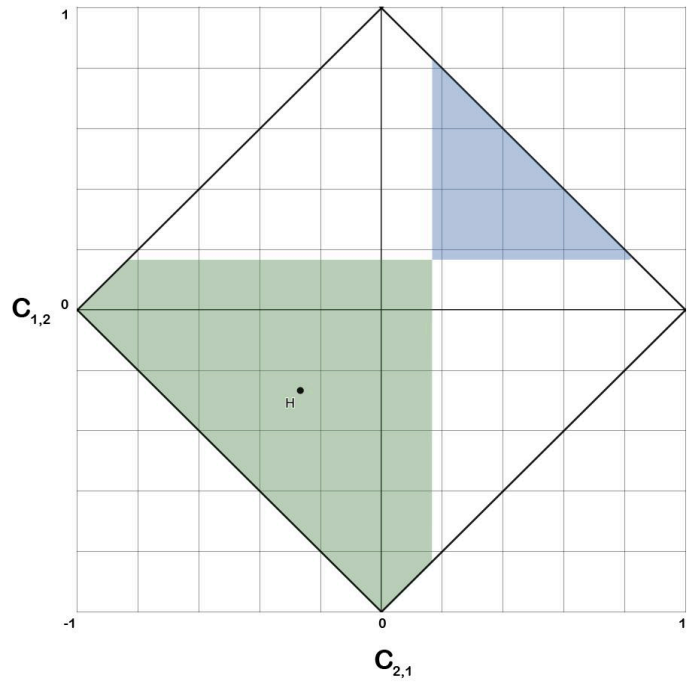

**Figure 5.** *Colonization space in Prisoner's Dilemma.*

The influence space $F_{DR}$ is instead a curved area. It is the region marked in red in *Figure 6*. $H' = (-\frac{4}{11}, -\frac{4}{11})$ is simply the transformation of $H$ from values of colonization to values of influence through [6] and [8].

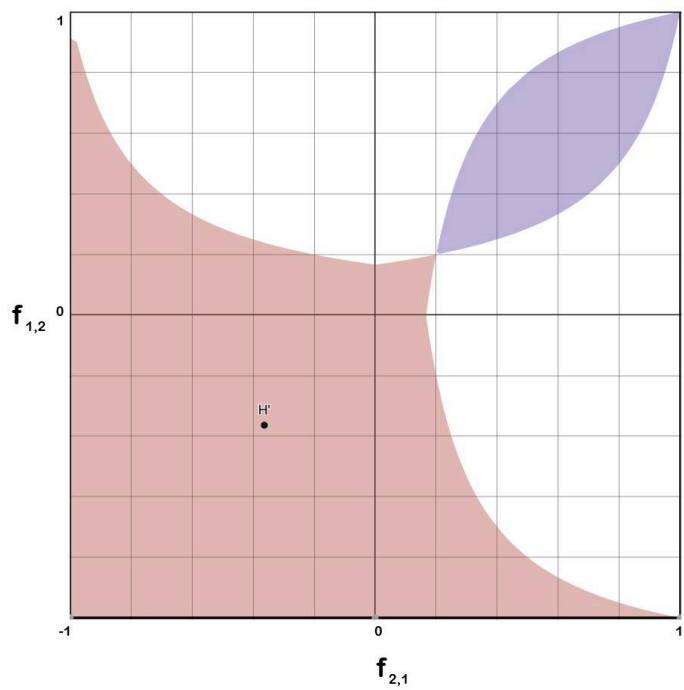

**Figure 6.** *Influence space in Prisoner's Dilemma.*

The plot shows that most of the σ area lies in the third quadrant. This suggests that the Nash equilibrium of our Prisoner's Dilemma can be symmetrically conflictual, because the third quadrant is where the influences of both players are negative: they are acting one against the other. $H$ and $H'$, the centroids of colonization and influence, also lie in the third quadrant.

The purple area in *Figure 6* is the influence space of $UL$, the profile where the two players cooperate, while the white areas of the square are the spaces of the two other profiles. In this particular case the four regions do not overlap. Overlap happens only in games where there are multiple Nash equilibria.

If the two players are neutral or in conflict, they will certainly defect. But if someone influences the other, or they influence each other, the equilibrium will change. Lastly, if we consider the distance from origin as a sort of energy of the system, we can say that the profile $UL$ requires more energy than the others.

As we did before, we can represent $H'$ as a graph and $H$ as a histogram. *Figure 7* depicts the centroid of influence of the Nash equilibrium in this Prisoner's Dilemma.

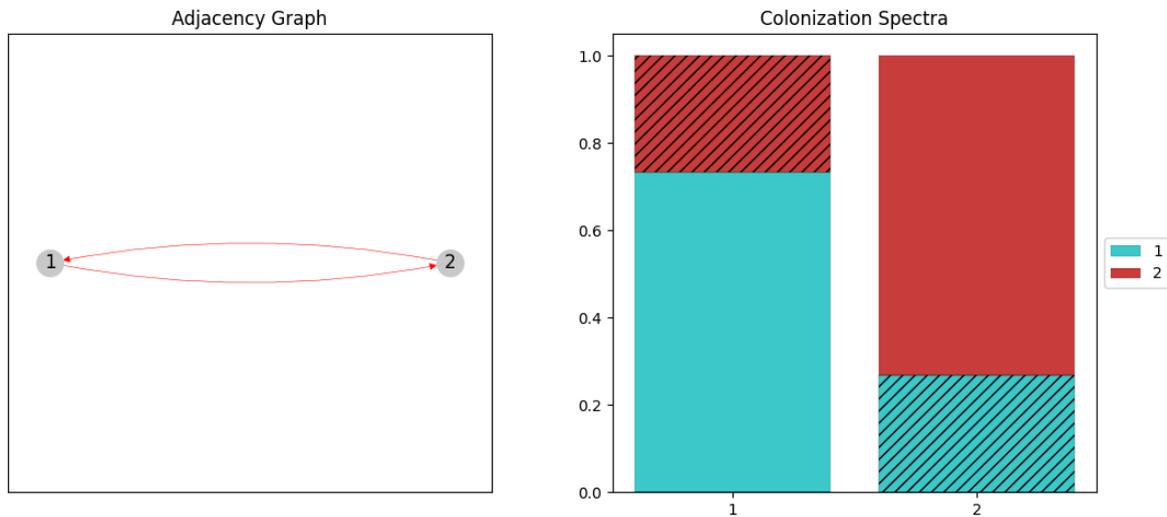

**Figure 7.** *Centroid of the Nash equilibrium in Prisoner's Dilemma.*

### *3.2 The Landowner Game*

The classic Cournot model concerns two or more firms competing. The firms decide how much product to bring on the market; if they compete, their profits decrease in favor of a third party, the consumer, who can buy a greater quantity of the commodity at a lower price. What happens in the system when we insert the consumer as a player?

Let us change our perspective. Now the agents who compete are peasants, who offer work in exchange for a wage, and the only consumer is a landowner. It is therefore a *monopsony*, because there is only one consumer and several producers (of work) competing. We will call it the Landowner Game.

Peasants decide how much to work for the landowner. Their hourly wage $W$ is set by this (inverse) demand curve:

**D7.** $\quad W = a - Q$

In this formula, $a$ (which in the examples below will be equal to 20) is the maximum amount of hours of work that the landowner would request if the peasants worked for free. While $Q = \sum_{i=1}^{n} q_i$ is the amount of work hours supplied by the peasants.

Pure utilities of workers are:

**D8.** $\quad u_i = (W - C)q_i$

where $C$ (cost) is the minimum wage a peasant could accept for work and will be 1 in the following examples. It should be noted that, as with firms, it is possible that $u_i = 0$ and $q_i > 0$. In other words, the peasants could also work with zero profit because in this case $W = C$: the wage is the least that peasants could accept.

But what is the utility of the landowner? In fact, the landowner is interested in having more work for less money. Yet the demand curve relates labor to wages in such a way that the more labor, the lower the wage, so ultimately the landowner's utility can be set as simply proportional to the amount of work:

**D9.** $\quad u_L = Q = a - W$.

Therefore, in the graph, the consumer node has a different utility function than the other nodes. Furthermore, in this simple model, the landowner cannot make any decision.

A node that cannot make decisions is passive, but its preferences can affect active nodes. In this case only the peasants, the active nodes, can make a choice: how much work to offer to the landowner, the passive node. But their choice also depends on the power the landowner has over them.

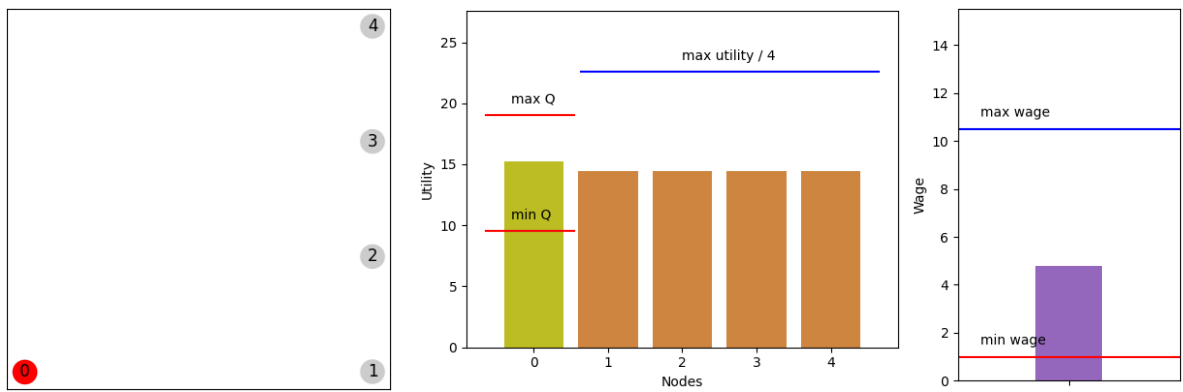

**Figure 8.** *Landowner Game, free system. Node zero (red) is the landowner: his/her utility (the olive green bar) can oscillate between max Q (perfect competition) and min Q (monopoly).*

In a free system of four peasants and a landowner (see *Figure 8*) we obviously have the same sub-optimal result as for the classic Cournot model. Workers compete with each other: each one offers more work for less money. But the limited number of workers prevents perfect competition, so neither the landowner reaches maximum utility.
The wage is well below the "max wage" line.

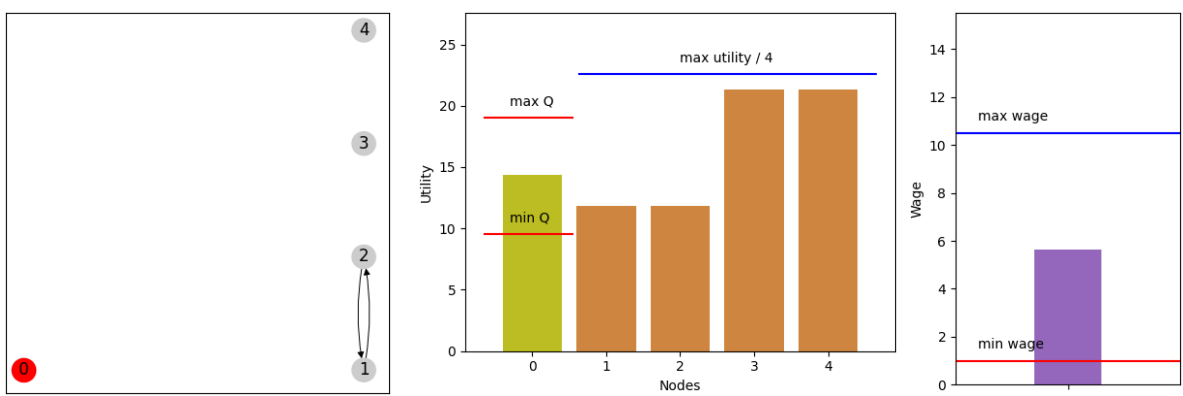

**Figure 9.** *Landowner Game, mutualism of only two nodes. If the union does not control all the farmers, then those who join lose. The wage is higher than in the free system and the quantity of work is smaller (therefore also the utility of the landowner is lower) (edges weight 0.8).*

Now suppose that a union is started (*Figure 9*), but that this organization includes only two individuals out of four. Quite surprisingly, those who join lose.
Imagine the workers striking to obtain a wage increase. If only two go on strike, wages increase because the quantity of work decreases, but all the profits go to the workers who do not strike, therefore they act as free riders.
In $F$-games, superadditivity and monotonicity fail: larger coalitions can gain less than the sum of their components.

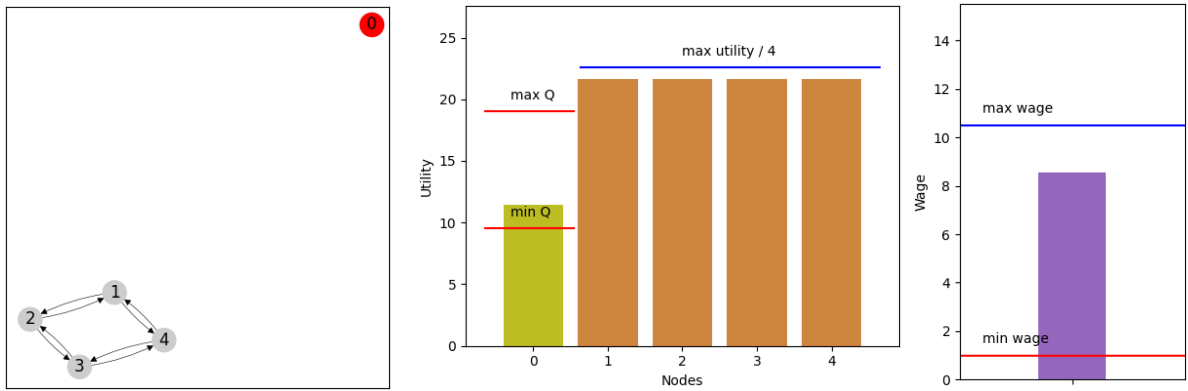

**Figure 10.** *Landowner Game, mutualism of all farmers. The union now has a monopoly on labor supply; the salary is maximum (edges weight 0.4).*

If, on the other hand, all the nodes join the union, as in *Figure 10*, the wage would reach its maximum and the amount of work would be minimal. In the middle graph, the olive bar indicating the amount of labor and the landlord's utility is near its minimum ("min Q" in *Figure 13b*), because the union now monopolizes the labor supply.

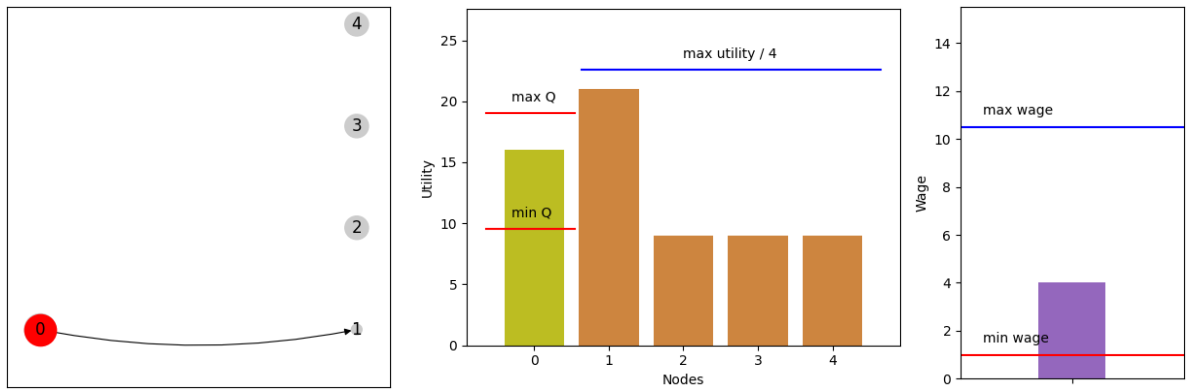

**Figure 11.** *Landowner Game, the landowner controls a farmer. The wage is lower than in the free system. The submissive peasant earns more than the others (edge weight 0.8).*

Suppose now that a peasant submits to the landowner. In doing so, the peasant would work more, making everyone's wages go down. But all in all, despite the drop in wages, the submissive peasant wins, as shown in *Figure 11*.

We are faced with a counterintuitive case, where submitting to someone implies an increase in pure utility for the submitted. Here obedience seems opportune.

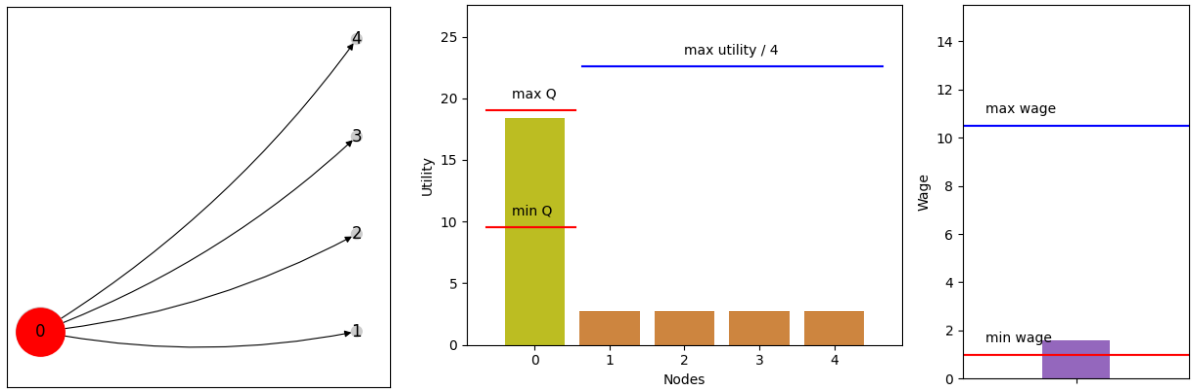

**Figure 12.** *Landowner Game, hierarchy dominated by the landowner. The wage is close to that of the perfect competition (edges weight 0.8).*

Nevertheless behind this apparent advantage there lies a trap for the peasants. In fact, if everyone acts like node (1) and submits to the landowner, the gain of (1) evaporates. All workers will be poorer, as in *Figure 12*. It is a Prisoner's Dilemma situation: individually, it is better for each farmer to submit, but the submission of everyone worsens the conditions for all. In this case the wage is close to the minimum that farmers need to cover the cost of an hour of work. On the other hand, the amount of work (represented by the green column) is the maximum.

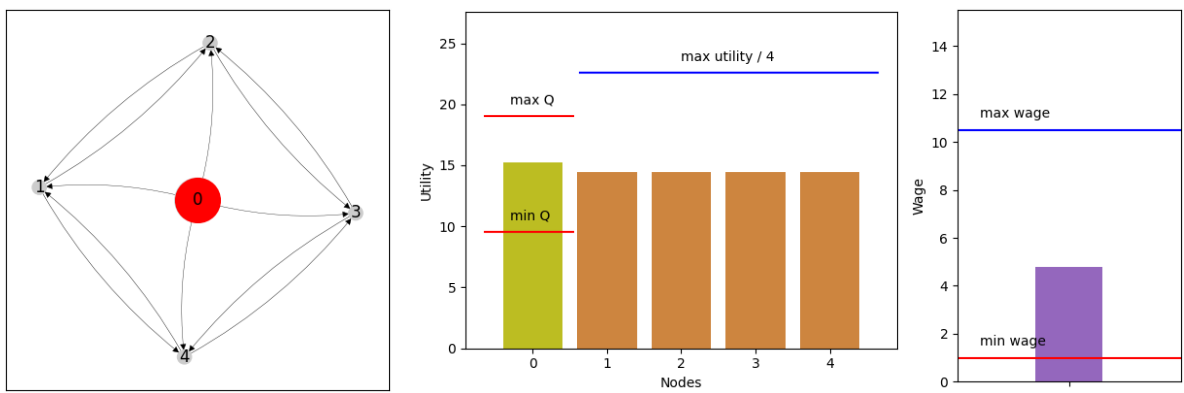

**Figure 13.** *Landowner Game, two powers that collide: a union of the peasants and a dominion of the landowner. The resulting equilibrium is the same as in the free system.*

Finally, the landowner can counterbalance the power of the union (*Figure 13*). The power of the union and that of the landowner cancel each other out as two equal and opposite forces. From the point of view of power structures, faced with a union of peasants, the landowner has two options: break the union link or increase control over the workers.

The Landowner Game shows an important property of $F$-games: the effect of power depends on the relationships among the pure preferences of the player. When two dominant players have pure preferences that cancel each other, a servant of two masters can act as a free player. That could also happen when the dominant player is in turn dominated by another player with opposite pure preferences. In some cases, the existence of a chain can consequently improve the servant's condition.

**4.1 The Lutheran Game**

We have seen so far how to confront different $F$-equilibria. With $F$-games we can measure actual power in an outcome of a game, as we did when we compute the influence spaces of strategy profiles on Prisoner's Dilemma. With $F$-games we can also confront the equilibria of two or more actual power networks, as we did for the Landowner Game.

The next $F$-game will clarify why we need to postulate *potential power* as something different from *actual power*, and why we need another kind of measure for it. I call it the Lutheran Game.

There are two players, God and man. Man wants salvation, but the choice is entirely in the hands of God.

**LUTHERAN GAME**     God
(*Table 2*)

|     | L | R |
|---|---|---|
| Man | -100 , 0 | 100 , 0 |

Man has no choice, because there is no action a creature could achieve to change the outcome. In this situation, actual power alone can say very little. Furthermore, any actual power of God over Man will be ineffective, because it would not change the fate of Man. The only effective influence is from Man to God: if God has positive sympathy over Man, there is salvation, if God's sympathy over Man is negative, damnation.

But it is impossible not to see that if there is any power in this game, it is all on God's side. God has no cost in choosing L or R, so that Man cannot predict what will occur.

So the only effective influence goes in the opposite direction of the power we naturally see in this game. We are facing the concept of *potential power*.

*4.2 Definition of potential power*

The Lutheran Game suggests a precise definition for potential power.

DEFINITION: A has as much *potential power* over B as A's sympathy for B could change the welfare of B.

While actual power deals with single profiles of strategies, potential power depends on the whole game. It has no negative side, because, by definition, the function that takes as input the sympathy of A for B and returns the welfare of B is always monotonously increasing. And it is not transitive, because potential powers of players are independent from each other.

I call this welfare function π. The value of $\pi_j(f_{j,i})$ is the mean value of the welfare of $j$ in all the equilibria in mixed strategies reached by the $F$-game when every entry of the $F$ matrix different from $f_{j,i}$ is null. The Nash theorem grant the existence of at least one mixed strategy equilibrium in every game, so $\pi_j(f_{j,i})$ is always defined for $f_{j,i} \in (-1, 1)$. We normalize π in order to obtain a function that returns 0 with $f_{j,i}$ null.

**D10.** $\overline{\pi}_j(f_{j,i}) = \pi_j(f_{j,i}) - \pi_j(0)$

We then obtain the potential power of $i$ over $j$ with the following integral:

**D11.** $P_{i,j} = \int_{-1}^{1} \left| \overline{\pi}_j(f_{j,i}) \right| df_{j,i}$

Eventually, if we want to confront potential power in different games, we can normalize $P_{i,j}$ as follows:

**D12.** $\overline{P}_{i,j} = \frac{P_{i,j}}{max(u_j) - min(u_j)}$

Potential power is in fact the capacity to reward or punish someone. It refers to what it could be: unlike actual power, it leaves the game unchanged. But, if B undergoes potential power from A, the same fact that B knows that can act as a promise or threat. The consequence is that B can consider rational to act differently, thus transforming potential in actual power. More precisely, B expects that A will use power for A's interest, either rewarding or punishing B if B either does or does not accomplish A's desires. So B can favor A with the intention of being rewarded.

But if B can favor A, B has potential power over A. So, the necessary condition for transforming potential into actual power is that there is some reciprocity in potential power.

This reveals something unexpected: the behavior of submission and of domination are, within the present perspective, the same behavior. A submits to or dominates B to achieve the same goal: being favored. In both cases A tries to change B's preferences, altering the feasible equilibria: A will favor when B favors, and harm when B harms. B will do the same. This means to adopt some kind of *trigger strategy*. In general, the idea is to match the sympathy of the other to the extent that $f_{j,i} = f_{i,j}$. Along this diagonal line the two welfare functions $\pi_i$ and $\pi_j$ are both monotonously increasing, so both players $i$ and $j$ prefer higher values of $f_{j,i} = f_{i,j}$. In concrete, this means to adopt different strategies that most of the time include some sympathy matching point and some peaceful first offer point.

Yet, there is no guarantee that all these strategies will be profitable. The transformation from potential to actual is not necessary, nor easily predictable. Rewarding or punishing who has potential power over us can be certainly said to be rational, but it is impossible to demonstrate that it is the best response in general. To be so, one had to be sure that most people behaved in the same way, rewarding and punishing the others. But one cannot easily predict if someone that has been punished will obey in the future. And one cannot be sure that helping someone very powerful will produce a gain in welfare.

They can only say something about the expected probability that a node adopts a trigger strategy with the intention of actualising its potential power: for instance, the more power it has, the less costly and more effective its action could be.

### 4.3 Examples of potential power

Let us apply the potential power formula to the Lutheran Game. Man cannot have any potential power, for two reasons: God is totally indifferent, and Man has no choice. The $\pi_M$ function is easy to compute.

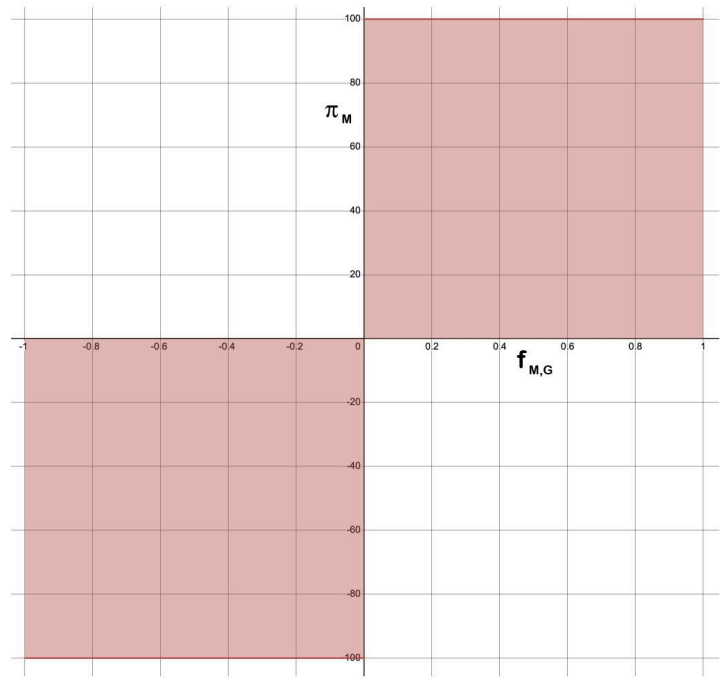

**Figure 14.** *The function $\overline{\pi}_M(f_{M,G})$. The red areas represent the value of $P_{G,M}$, the potential power of God over Man.*

- When $f_{M,G} = 0$ God is totally indifferent between *L* and *R*, so there are infinite mixed equilibria. The mean value of Man's payoff is 0.
- When $f_{M,G} > 0$, God chooses *R* and Man obtains a payoff of 100.
- When $f_{M,G} < 0$, God chooses *L* and Man obtains a payoff of -100.

As we see in *Figure 14*, the areas of the two rectangles are 100, so $P_{G,M} = 200$ and $\overline{P_{G,M}} = 1$. We can say that God has total potential power over Man.

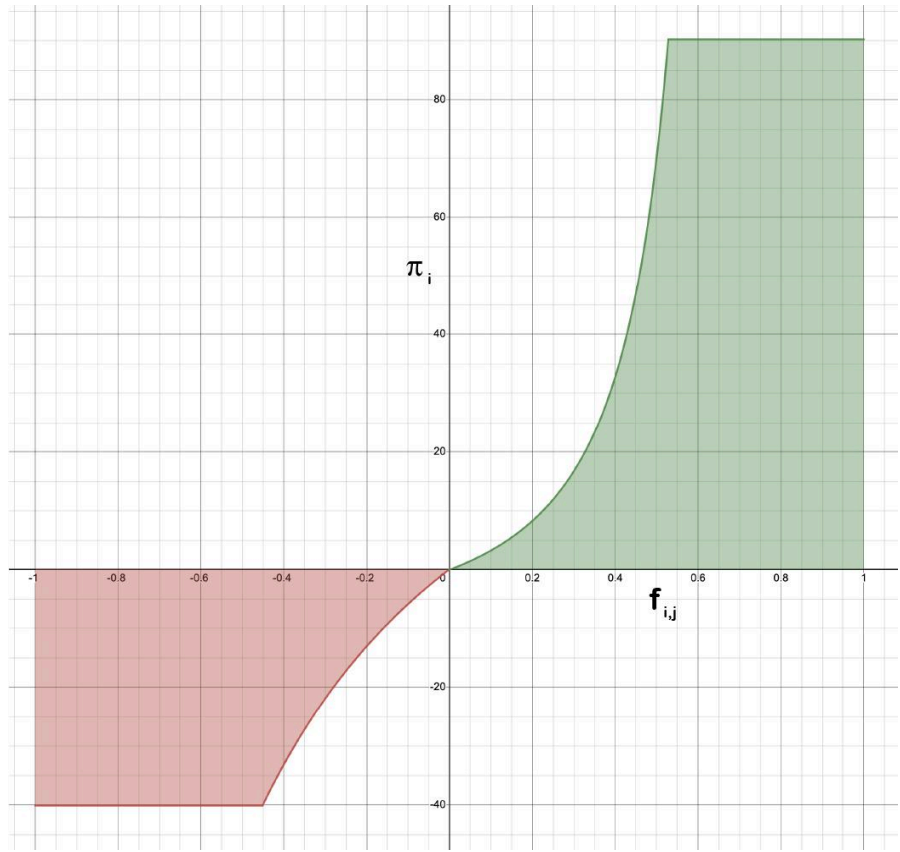

**Figure 15.** *The function $\overline{\pi}_i(f_{i,j})$ in the Cournot Model and in the Landowner Game (with a=20, C=1). The red and green areas represent the value of $P_{j,i}$, the potential power of a firm over another firm, or of a peasant over another peasant.*

*Figure 15* depicts the potential power from peasant to peasant in a three players (two peasants and a landowner) Landowner Game. Power is the sum of the red and green areas. Landowner has no power over peasants because, in our simplified game, the landowner's node is passive.

We see that the green area is bigger than the red area. As well as in other games similar to Prisoner's Dilemma, the influence space of the Nash equilibrium with $F$ null is conflictual: 0 is nearer to the bottom of the plot. In terms of potential power, this means more possibilities on the cooperative side. This is consistent with the consideration we made about the influence space of the Nash equilibrium in Prisoner's Dilemma.

### *5.1 Conclusions and further questions*

We have seen so far how, by mixing the preferences of the players, we can simulate situations of submission or domination, coalition or mutual aid, as well as of harm or punishment. We saw several examples of applying power structure in the Landowner Game, along with their consequences on equilibria. Furthermore, we were able to compute the influence space of a given strategy profile, i.e. the set of the power structures that allows to reach that profile as an equilibrium.

We recognized these mixed preferences as actual power, meaning that it actually changes the payoff (and by consequence, the actions) of the players.

We then realize the necessity of the notion of potential power, that we defined as the integral of the function that takes as input the sympathy of a player and returns the mean payoff of another. While actual power deals with favoring and harming, potential power deals with promising and threatening. Potential power does not alter any welfare, but it may induce a player to adopt some kind of *trigger strategy* in order to minimize risk. This suggests one possible origin of actual power: for example, it could be rational for a player to favor another if the latter is threatening the former.

The last example shows how potential and actual power are, under a certain respect, reversed: threatening and favoring tend to go in opposite directions.

Yet, there remain several questions about the relation between actual and potential power. How can potential power transform to actual, so becoming transitive and relative? Is it possible a perfect transformation? How $F$-games relate to other traditional indices of power in games?

These and other questions go beyond the intentions of this paper.

*Bibliography*

Allingham, M. G. (1975). Economic power and values of games. *Zeitschrift für Nationalökonomie/Journal of Economics*, (H. 3/4), 293-299.

Banzhaf III, J. F. (1964). Weighted voting doesn't work: A mathematical analysis. *Rutgers L. Rev.*, *19*, 317.

Bardhan, P. (1991). On the concept of power in economics. *Economics & Politics*, *3*(3), 265-277.

Bartlett, R. (2006). Economics and power. *Cambridge Books*.

Binmore, K. (2009). Interpersonal comparison of utility.

Binmore, K. (2010). Social norms or social preferences?. *Mind & Society*, *9*, 139-157.

Boella, G., Sauro, L., & van der Torre, L. (2005, September). Admissible agreements among goal-directed agents. In *IEEE/WIC/ACM International Conference on Intelligent Agent Technology* (pp. 543-549). IEEE.

Boella, G., Sauro, L., & van der Torre, L. (2006, May). Strengthening admissible coalitions. In *ECAI* (Vol. 6, pp. 195-199).

Bowles, S., Franzini, M., & Pagano, U. (Eds.). (1998). *The politics and economics of power*. Routledge.

Bowles, S., & Gintis, H. (2011). *A Cooperative Species*. Princeton University Press.

Castelfranchi, C. (1998). Modelling social action for AI agents. *Artificial intelligence*, *103*(1-2), 157-182.

Castelfranchi, C., Miceli, M., & Cesta, A. (1992). Dependence relations among autonomous agents. *Decentralized AI*, *3*, 215-227.

Dahl, R. A. (1957). The concept of power. *Behavioral science*, *2*(3), 201-215.

Fehr, E., & Charness, G. (2023). Social preferences: fundamental characteristics and economic consequences.

Grossi, D., & Turrini, P. (2010, May). Dependence theory via game theory. In *AAMAS* (pp. 1147-1154).

Grossi, D., & Turrini, P. (2012). Dependence in games and dependence games. *Autonomous Agents and Multi-Agent Systems*, *25*, 284-312.

List, J. A. (2009). Social preferences: Some thoughts from the field. *Annu. Rev. Econ.*, *1*(1), 563-579.

Matteucci, N., Pasquino, G., & Bobbio, N. (Eds.). (1991). *Dizionario di politica*. Editori associati.

Shapley, L. S., & Shubik, M. (1954). A method for evaluating the distribution of power in a committee system. *American political science review*, *48*(3), 787-792.

Sichman, J. S. (1998). Depint: Dependence-based coalition formation in an open multi-agent scenario. *Journal of Artificial Societies and Social Simulation*, *1*(2), 1998.

Taylor, M. (1982). *Community, anarchy and liberty*. Cambridge University Press.

Wiese, H. (2009). Applying cooperative game theory to power relations. *Quality & quantity*, *43*, 519-533.